\documentclass[dvips]{article}
\usepackage{icrctc07}

\title{Calibration systems and methods for the ANTARES neutrino telescope}
\shorttitle{Calibration of the ANTARES neutrino telescope}

\authors{F. Fehr$^{1}$ on behalf of the ANTARES collaboration$^{2}$}
\shortauthors{F. Fehr et al.}
\afiliations{$^1$Univ. Erlangen-Nuremberg, Physics Institute 1, Erwin-Rommel-Str.1, 91058 Erlangen, Germany\\ 
             $^2$http://antares.in2p3.fr }
\email{fehr@physik.uni-erlangen.de}

\abstract{The ANTARES neutrino telescope is currently being constructed in the Mediterranean Sea.
The complete detector will consist of 12~strings, supplemented by an additional instrumentation line. 
Nine strings are at present deployed of which five are already connected to the shore and operating.
Each string is equipped with 75~Optical Modules (OMs) housing the photomultipliers
to detect the Cherenkov light induced  by the charged particles produced in
neutrino reactions. 

An accurate measurement of the Cherenkov photon arrival times as well as the positions and
orientations of the OMs is required for a precise reconstruction of the direction of the 
detected neutrinos. For this purpose the ANTARES detector is provided with several systems to facilitate
the calibration of the detector. The time calibration is performed
using light pulses emitted from LED and laser devices.
The positioning is done via acoustic triangulation using hydrophones. Additionally, local tilt angles
and the orientations of the modules are measured with a set of tiltmeters and compasses.

In this paper, it is demonstrated that the ANTARES time and alignment calibration
systems operate successfully in situ. In particular, it is shown that the ANTARES read-out electronics is capable 
of reaching a sub-nanosecond time resolution.}

\begin{document}
\maketitle
\section{Introduction}
The ANTARES neutrino telescope \cite{ref0:status} is being constructed at a depth of 2.5~km 
in the Mediterranean Sea 40~km off shore from Toulon (France).

The complete detector will consist of 12~strings (also referred to as lines) of 450~m length with about 70~m spacing, 
anchored on the sea bed and maintained vertically by buoys.
Each string is equipped with 75~Optical Modules \cite{ref1:ompaper},
pressure resistant glass spheres housing a $10"$ photomultiplier tube (PMT), which detects the Cherenkov 
light emitted by charged particles produced in neutrino reactions with the surrounding matter.
The OMs of a string are arranged into triplets on 25~storeys of 14.5~m vertical distance to each other, with the PMTs looking 
downwards at an angle of $45^\circ$ with respect to the horizontal.

In 2006, the first ANTARES lines have been deployed and put into operation \cite{ref2:firstmuons}.
Presently, nine strings have been deployed of which five are already connected to shore and taking data continuously.
The completion of the ANTARES telescope is foreseen for the beginning of 2008.

The high angular resolution for neutrino astronomy ANTARES aims at ($0.3^\circ$ for muon events above 10~TeV)
is intimately connected with a precise resolution of the relative arrival times of Cherenkov photons at the PMTs.
The relative time resolution between OMs is therefore of importance. The time resolution is intrinsically limited
by the transit time spread of the signal in the PMTs ($\sigma \approx 1.3\,\rm{ns}$) and the optical properties of the seawater
\cite{ref3:opticalprops} such as light scattering and chromatic dispersion ($\sigma \approx 1.5\,\rm{ns}$ at a distance of 40 meters). 

The ANTARES read-out electronics and calibrations systems are designed to contribute less than 0.5~ns to the overall timing resolution.
In addition, a precise measurement of the relative positions of the OMs with an accuracy of 10-20~cm is required.

\section{Time calibration}
The time calibration of the ANTARES telescope is performed using \textit{several systems} which provide complementary 
information on the propagation of the signal in the detector.

\textbf{The internal clock calibration system.} A precise time reference clock system
has been implemented in the ANTARES detector. This system consists of an on shore 20~MHz clock generator connected via a clock
distribution system to the clock signal transceiver boards integrated in each storey. 

This setup allows to measure the time offsets of a storey using an echo-based time calibration whereby each storey sends
back a return signal through the same optical path as the outgoing clock signal. In situ measurements show an excellent 
stability of $\approx 0.1\,\rm{ns}$.  Furthermore the internal clock system facilitates an absolute event time measurement, 
with a precision of about $100\,\rm{\mu s}$, by assigning a GPS timestamp to the data. This absolute precision is more
than sufficient to attribute possible astrophysical sources to the recorded neutrino events.

\textbf{The internal Optical Module LEDs.} A blue Agilent HLMP-CB15 LED is mounted on the back of the PMT inside the 
Optical Modules. The light intensity of the LEDs peaks at about 470~nm with a FWHM of 15~nm. These LEDs are used to 
measure the relative variation of the PMT transit time.

\textbf{The Optical Beacons.} The relative time calibration including 
the optical properties of the sea water at the ANTARES site is accomplished using LED and Laser devices, so called
Beacons \cite{ref4:beaconpaper}.

Four LED Beacons are distributed along every string.
Each LED Beacon contains 36 individual
LEDs.
The Beacon LEDs have been synchronized in the laboratory to better than 0.1~ns. The light pulse emitted by the LED Beacons
has a rise time of about 2.8~ns.

The Laser Beacon is a more powerful device located at the bottom of selected strings. It uses a diode pumped Q-switched Nd-YAG laser to produce pulses of $\approx 1\,\rm{\mu J}$ with a FWHM of $\approx 0.8\,\rm{ns}$ at a wavelength of 532~nm. The laser beam is widened by a diffuser which spreads the light out according to a cosine angular distribution. This permits to illuminate adjacent lines. 

\subsection{Dark room time calibration}
Prior to deployment, the strings are tested and pre-calibrated in the laboratory using a dark room
setup. An attenuated optical signal pulse is sent from a Nd-YAG solid state laser via an optical splitter to the OMs. One output of the splitter is connected to a reference control module that is used to relate the timing of the different OMs. The dark room calibration provides a set of calibration values for first in situ data taking and is used to cross check subsequent in situ calibrations.  Additionally, a test of the clock system is performed in the dark room.

\subsection{In situ time calibration}
In situ time calibrations have been performed regularly using dedicated measurement setups for all of the five operating lines. 

\begin{figure*}[t!]
\begin{center}
\noindent 
\includegraphics [width=0.99\textwidth]{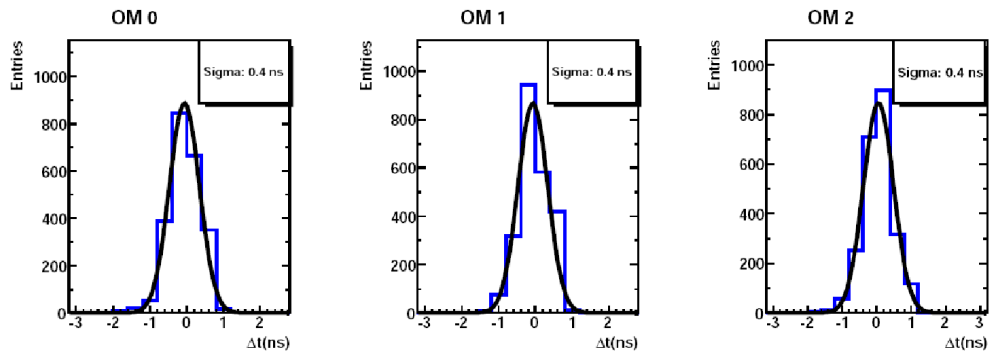}
\end{center}
\caption{Distributions of the difference between the time of arrival of the  light at the three OMs of storey~3 of Line~1 and the time of emission from the LED Beacon in storey~2 on the same line. The means have been set to zero.
\cite{ref4:beaconpaper}}\label{fig_led_beacon_calib}
\end{figure*}

Figure \ref{fig_led_beacon_calib} shows, as an example, the time difference distributions between the time of arrival of the light at the three OMs of storey~3 of Line~1 and the time of emission from the LED beacon in storey~2 on the same line. Due to the short distance between the beacon and the OMs ($\approx 13.4\,\rm{m}$) and the high light intensity of the beacon pulse, contributions from the transit time spread in the PMT, from the signal pulse walk, as well as light scattering and line movements are negligible. Hence, the width of the
time distribution ($\sigma \approx 0.4\,\rm{ns}$) reflects the resolution of the read-out electronics and is in agreement with the expected time resolution \cite{ref5:arspaper}.

\section{Alignment of the detector}
Being exposed to the sea current, the ANTARES strings undergo drifts of several meters.
Therefore, a second essential element of the calibration is the real time measurement of the OM positions within an accuracy of 10-20 cm.
Again, \textit{several complementary systems} are used to measure the positions of the detector elements.

\textbf{Acoustic modules.} The acoustic modules \cite{ref6:acousticpos} transduce electrical signals to sound waves and vice versa. The 
transducing part made of a hemispherical piezo-ceramic (1~cm diameter) is referred to as hydrophone.
The acoustic modules comprise emitter-receivers (RxTx) located at the anchors of ANTARES
strings, receivers (Rx) located on the lines and transponders. 
Transponders are autonomous units located on the sea bed around the lines at some 300~m distance
from each other, providing additional triangulation points to increase the precision of the
measurement.

\textbf{Oceanographic probes.} Beside the acoustic emitters and receivers, additional
instruments to measure the sound velocity at the site are needed to translate the propagation 
time of the acoustic signals between two modules into their distance, thus obtaining a relative positioning of the detector. The sound velocity depends on three main properties of the sea water: salinity, temperature and pressure. 
In order to measure the sound speed variations within the detector, ANTARES utilizes a set of sound velocimeters, conductivity, temperature
and pressure probes coupled to sound velocimeters, as well as separate, independent pressure probes.

\textbf{Compasses and tiltmeter systems.} A system of compasses and tiltmeters on every storey of the strings
supplements the acoustic positioning. The compasses measure the orientation of the storeys and the tiltmeters
provide the inclination of the string.

\subsection{In situ acoustic positioning}

\textbf{Relative positioning.} The acoustic triangulation is carried out with a period of a few minutes. 
Triggered by the ANTARES clock system, the RxTx acoustic modules alternately send
signals to the other modules being in listening mode. In addition, the triangulation using 
the transponders is processed separately.

\begin{figure}[t!]
\begin{center}
\includegraphics [width=0.5\textwidth]{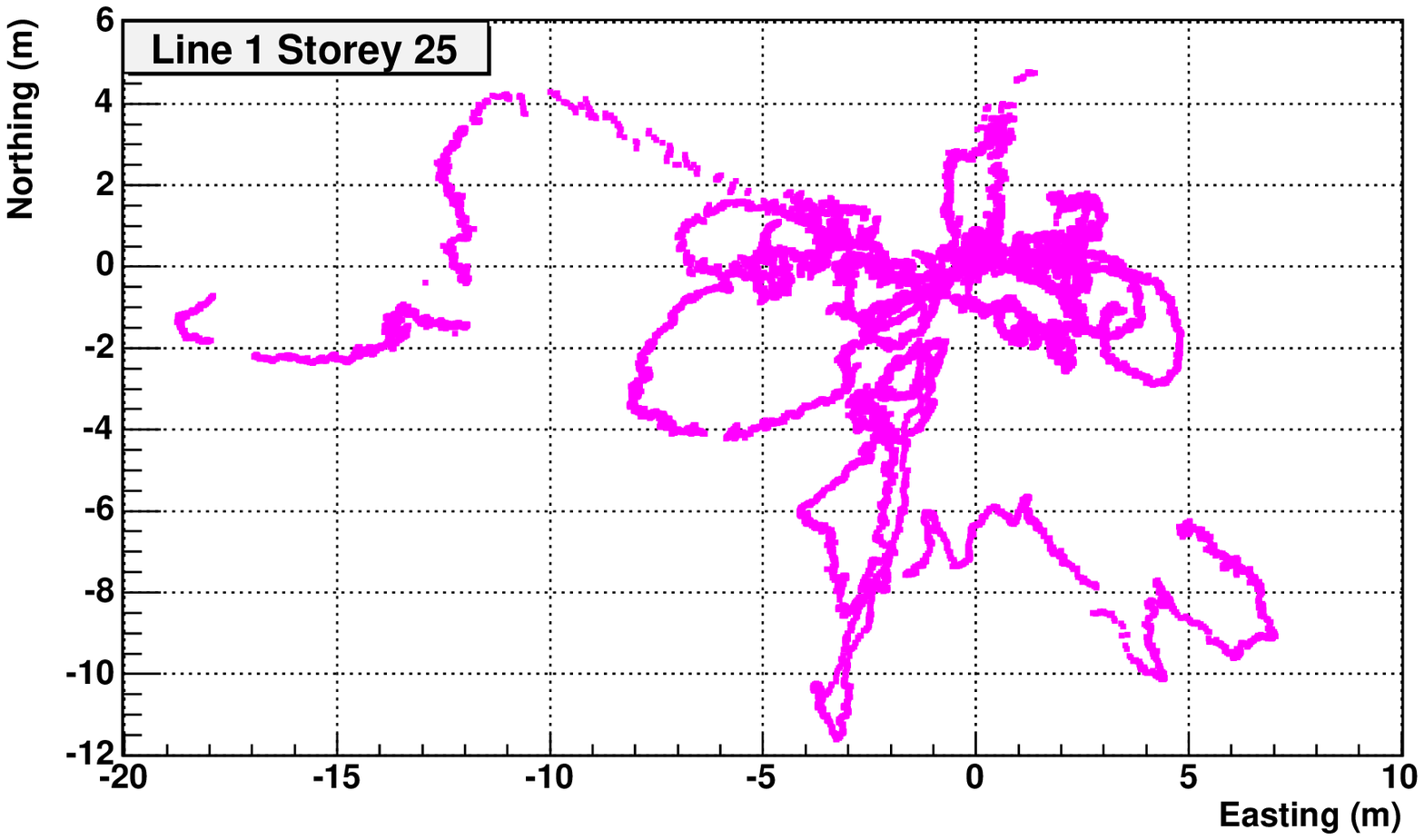}
\includegraphics [width=0.5\textwidth]{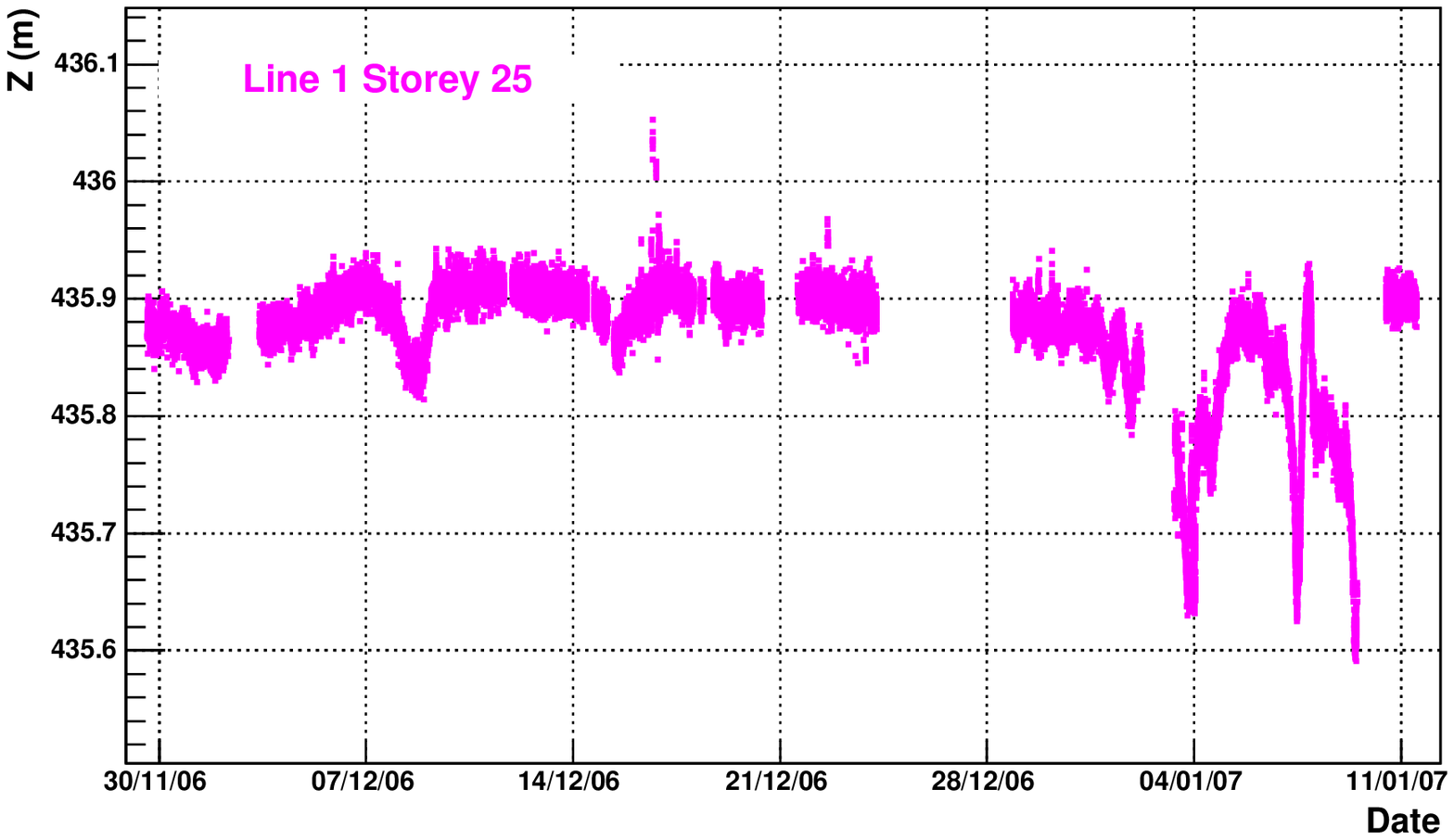}
\end{center}
\caption{\textbf{(Top)} Horizontal trajectory of the hydrophone located at the top storey of Line~1
from the end of Nov'06 to mid Jan'07. As expected from the main current direction on the 
ANTARES site, the largest displacement of the line is observed towards West. \textbf{(Bottom)}
Vertical displacements due to the inclination of the line of the same hydrophone in Dec'06-Jan'07.}\label{fig_ac_pos}
\end{figure}

An example of the measurements is given in Fig.~\ref{fig_ac_pos}. Here, the horizontal trajectory as
well as the vertical displacements of the hydrophone on top of Line~1 for a certain data period are shown,
see Figure caption for further details. The accuracy of the measurement is well within the specifications.

\textbf{Absolute positioning.} The absolute positioning of the ANTARES telescope has been performed by acoustic
triangulation of the detector elements from a surface ship equipped with a differential GPS antenna. 
The triangulation with the ship has been repeated at different ship positions measured by the
DGPS receiver. Accuracies of about 1~m are achieved on geodetic positions of anchored detector components.

\subsection{Line shape model}
\begin{figure}[t!]
\begin{center}
\noindent
\includegraphics [width=0.48\textwidth]{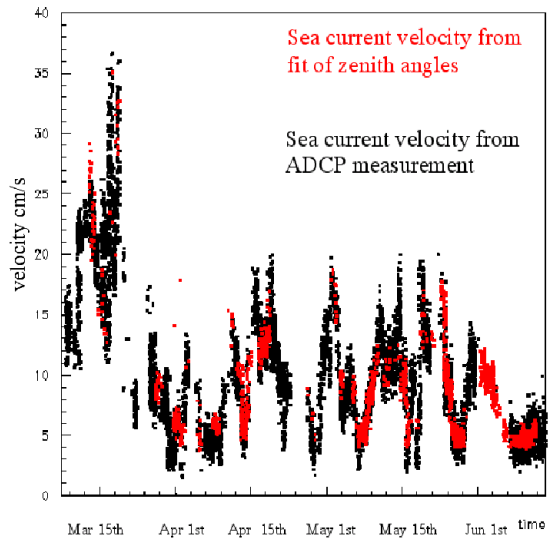}
\end{center}
\caption{Sea current velocity measured with the MILOM ADCP compared to the result of 
the line shape model fit for Line~1 data from March to June~2006}\label{fig_sea_current}
\end{figure}

The movements of the strings are caused by the sea current. A model has been developed which allows 
to calculate the line shape based on the drag forces due to the sea current and the known buoyancy of the detector elements. 
The free parameters in this model are the sea current velocity and direction which can be fitted to the data.
A comparison (see Fig.\ref{fig_sea_current}) between the sea current velocity obtained by the model and
the measurement with an Acoustic Doppler Current Profiler (ADCP) on the MILOM test line shows a good
agreement for a wide range (5-35~cm/s) of sea current velocities.

\section{Conclusions}
In situ calibration for all five operating strings have been performed regularly. The time calibration
as well as the positioning systems work well within the specifications. 

It has been demonstrated that the ANTARES read-out electronics is capable of reaching a sub-nanosecond time resolution in situ.

\textit{This work is supported by the German BMBF Grant No. 05 CN5WE1/7.}

\bibliography{icrc0481}

\begin{thebibliography}{1}

\bibitem{ref0:status}
A.~Kouchner, ``{The ANTARES neutrino telescope: status report},'' {\em In this
  proceedings}, 2007.

\bibitem{ref1:ompaper}
P.~Amram {\em et~al.}, ``{The ANTARES optical module},'' {\em Nucl. Instrum.
  Meth.}, vol.~A484, pp.~369--383, 2002.

\bibitem{ref2:firstmuons}
G.~Giacomelli and P.~Kooijman, ``{ANTARES Collaboration detects its first
  muons},'' {\em CERN Cour.}, vol.~46N7, pp.~24--25, 2006.

\bibitem{ref3:opticalprops}
J.~A. Aguilar {\em et~al.}, ``{Transmission of light in deep sea water at the
  site of the ANTARES neutrino telescope},'' {\em Astropart. Phys.}, vol.~23,
  pp.~131--155, 2005.

\bibitem{ref4:beaconpaper}
M.~Ageron, ``{The ANTARES optical beacon system},'' 2007.
\newblock Accepted for publication by Nucl.Instrum.Meth., astro-ph/0703355.

\bibitem{ref5:arspaper}
F.~Feinstein, ``{The analogue ring sampler: A front-end chip for ANTARES},''
  {\em Nucl. Instrum. Meth.}, vol.~A504, pp.~258--261, 2003.

\bibitem{ref6:acousticpos}
V.~Niess, ``{Underwater acoustic positioning in ANTARES},'' {\em {Proceedings
  of 29th ICRC, vol. 5}}, 2005.

\end{thebibliography}
\bibliographystyle{ieeetr}
\end{document}